\documentclass[structabstract]{raa}
\usepackage{graphicx,times}
\usepackage{natbib,amssymb}
\usepackage{epsfig}
\usepackage{float}

\providecommand{\bm}[1]{\mbox{\boldmath$#1$\unboldmath}}

\begin{document}
   \title{Binary population synthesis for the core-degenerate scenario of type Ia supernova progenitors$^*$
   \footnotetext{\small $^*$Supported by the National Natural Science Foundation of China.} }

   \volnopage{ {\bf 2015} Vol.\ {\bf X} No. {\bf XX}, 000--000}
   \setcounter{page}{1}

   \author{W.-H. Zhou \inst{1,2,3}
          \and
          B. Wang \inst{4,5}
          \and
          X. Meng \inst{4,5}
          \and
          D.-D. Liu \inst{2,4,5}
          \and
          G. Zhao \inst{1}
          }

   \institute{Key Laboratory of Optical Astronomy, National Astronomical Observatories, Chinese Academy of Sciences, Beijing 100012, China\\
          \and
          University of Chinese Academy of Sciences, Beijing 100049, China\\
          \and
          Yunnan Minzu University, Kunming 650031, China\\
          \and
          Yunnan Observatories, Chinese Academy of Sciences, Kunming 650026, China
          {\it wangbo@ynao.ac.cn; xiangcunmeng@ynao.ac.cn}\\
          \and
          Key Laboratory for the Structure and Evolution of Celestial Objects, Chinese Academy of Sciences, Kunming 650026, China
              }

   \date{Received ; accepted}

\abstract {The core-degenerate (CD) scenario has been suggested to be a possible progenitor model of type Ia supernovae (SNe Ia), in which a carbon-oxygen white dwarf (CO WD) merges with the hot CO core of a massive asymptotic giant branch (AGB) star during their common-envelope phase. However, the SN Ia birthrates for this scenario are still uncertain. We conducted a detailed investigation into the CD scenario and then gave the birthrates for this scenario using a detailed Monte Carlo binary population synthesis  approach. We found that the delay times of SNe Ia from this scenario are $\sim$70\,Myrs$-$1400\,Myrs, which means that the CD scenario contributes to young SN Ia populations. The Galactic SN Ia birthrates for this scenario are in the range of $\sim$7.4$\times10^{-5}\rm yr^{-1}$ $-$3.7$\times10^{-4}\rm yr^{-1}$, which roughly accounts for $\sim$2$-$10\% of all SNe Ia. This indicates that, under the assumptions made here, the CD scenario only contributes a small portion of all SNe Ia, which is not consistent with the results of Ilkov \& Soker (2013).
 }

\titlerunning{BPS for the CD scenario of SNe Ia}

\authorrunning{W.-H. Zhou et al.}

\maketitle

%

\section{Introduction} \label{1. Introduction}

Type Ia supernovae (SNe Ia) play an important role in astrophysics. Due to their high luminosities and uniformity, SNe Ia are considered to be  good distance indicators to determine cosmological parameters (e.g., Schmidt et al. 1998; Riess et al. 1998; Perlmutter et al. 1999). Studies on measuring cosmological distance through SNe Ia indicate that the expansion of the Universe is accelerating, which implies the existence of dark energy (see also, e.g., Riess et al. 2007; Kuznetsova et al. 2008).  In addition, SNe Ia are also crucial for the study of galactic chemical evolution as they are the main contributes of iron to their host galaxies (e.g., Greggio \& Renzini 1983; Matteucci \& Greggio 1986).

There is a theoretical consensus that SNe Ia are thermonuclear explosions of carbon-oxygen white dwarfs (CO WDs) in binary systems (Hoyle \& Fowler 1960; Nomoto et al. 1997). However, the precise nature of SNe Ia remains uncertain, especially concerning their progenitor models and explosion mechanisms (Leibundgut 2000; Hillebrandt \& Niemeyer 2000; Parthasarathy et al. 2007; Podsiadlowski et al. 2008; Bogomazov \& Tutukov 2011; Wang \& Han 2012; Wang et al. 2013b). The most remarkable SN Ia properties are their apparent similarity to each other. Most of the discussions about possible progenitors of SNe Ia are mainly concentrated upon the Chandrasekhar mass model. When the WD increases its mass to the Chandrasekhar mass limit, it may explode as an SN Ia.

Several SN Ia progenitor scenarios have been proposed over the past few decades, e.g., the single-degenerate (SD) scenario where the companion of the CO WD is a non-degenerate star (e.g., Whelan \& Iben 1973; Nomoto et al. 1984; Hachisu et al. 1996; Li \& van den Heuvel 1997; Han \& Podsiadlowski 2004; Meng et al. 2009, 2011; Chen \& Li 2009; Wang et al. 2009a, 2010, 2014b; Ablimit et al. 2014), the double-degenerate (DD) scenario where SNe Ia arise from the merging of two CO WDs (e.g., Tutukov \& Yongelson 1981; Webbink 1984; Iben \& Tutukov 1984), the double-detonation scenario where a sub-Chandrasekhar mass WD accumulates a layer of He-rich material from a He donor star (e.g., Woosley \& Weaver 1994; Livne \& Arnett 1995; Wang et al. 2013a), and the WD$-$WD collision scenario where two WDs collide and immediately ignite (e.g., Raskin et al. 2009; Rosswog et al. 2009; Kushnir et al. 2013). Each of the above scenarios is not completely consistent with observations at present. Observational evidence suggests that these scenarios may coexist (see the review by Howell 2011; Wang \& Han 2012; Maoz et al. 2014).

Early numerical simulations have shown that the deficiency of the DD scenario is its tendency to result in an accretion-induced collapse (AIC) and, ultimately, the formation of a neutron star (Saio \& Nomoto 1985; Hillebrandt \& Niemeyer 2000).\footnote {Even if the AIC can be avoided, the WD remnant of merger might lose about $0.5\,M_{\odot}$ via a super wind from the giant-like structure and fail to fulfil the critical mass for the SN explosion (e.g., Willson 2007; Shen et al. 2011).} In order to overcome the AIC of the DD scenario, the core-degenerate (CD) scenario has been proposed. In this scenario, a Chandrasekhar or super-Chandrasekhar mass WD is formed from the merger of a CO WD with the hot core of an AGB star during the planetary nebula phase or shortly after the termination of the common envelope (CE) phase (e.g., Sparks \& Stecher 1974; Livio \& Riess 2003; Kashi \& Soker 2011; Soker 2013a; Ilkov \& Soker 2012, 2013; see also Tout et al. 2008; Mennekens et al. 2010). Soker et al. (2014) suggested that the properties of SN 2011fe (e.g., the carbon rich composition of the fast moving ejector and a compact exploding object) may be explained by this scenario. Recently, Briggs et al. (2015) also did some population synthesis of WD$+$AGB core merger, and found that the majority of the high field magnetic WDs are the carbon-oxygen type and merge within a common envelope.

It was originally assumed that circumstellar material (CSM) would not be present in the merger of two WDs (e.g., Maguire et al. 2013). However, recent theoretical studies, which investigated the interaction of ejected material from the WDs with the interstellar medium (Raskin \& Kasen 2013; Shen et al. 2013), have suggested that the detectable CSM in some SNe Ia could be explained by the DD scenario (see  also Ruiter et al. 2013). Soker et al. (2013) argued that a prompt violent merger via the CD scenario can explain the properties of SN PTF 11kx, e.g., the multiple shells of CSM and the interaction of ejected material from SN with the CSM which started 59\,d after the explosion of the SN. Note that Dilday et al. (2012) suggested that SN PTF 11kx can be explained by the SD scenario.

Although the CD scenario has many advantages, as mentioned above, which might explain some properties of SN Ia diversity, many of the characteristics of the CD scenario obtained from different methods are not consistent with each other, especially the SN Ia birthrate. Ilkov \& Soker (2013) recently calculated the expected number of SNe Ia in the CD scenario, and their results showed that the CD scenario can account for the birthrates of SNe Ia within the uncertainties of several processes. The estimated SN Ia birthrate for this scenario is higher than that of observed, based on their simulations, assuming certain values for the parameters in their model. The purpose of this paper is to study SN Ia birthrates and delay times for this scenario using a detailed binary population synthesis (BPS) approach. In Section 2, we describe the BPS methods for the CD scenario. In Section 3, we show the simulation results of the CD scenario by the BPS approach. Finally, we present a discussion and conclusions in Section 4.

\section{Methods}

Adopting tested assumptions regarding the CD scenario (Soker 2013a; Ilkov \& Soker 2012, 2013), we performed a series of Monte Carlo simulations via Hurley's rapid binary evolution code (Hurley et al. 2000, 2002). In each simulation, we have followed the evolution of $1\times10^{\rm 7}$ sample binaries, some of which could form WD + AGB binaries. The criteria for potential SN Ia progenitors for the CD scenario are as follows. (1) The total mass of the WD remnant of the primary (${M}_{\rm WD}$) and the mass of the AGB core (${M}_{\rm core}$, secondary) during the final common envelop (CE) phase should be super-Chandrasekhar, i.e., ${M}_{\rm WD}+{M}_{\rm core} \geqslant 1.4M_{\odot}$. (2) ${M}_{\rm core} \geqslant {M}_{\rm WD}$, the core of the AGB star has a greater mass than the WD remnant of the primary star. (3) The WD and the AGB core merge in the subsequent CE phase.

The binary formation channel for the CD scenario in this paper is similar to that described in Ilkov \& Soker (2013).
The mass of the primordial primary star is in the range of $2.0$-$6.5\,M_{\odot}$, and the initial mass ratio between the secondary and the primary $(M_{2}/M_{1})$ is in the range of 0.76$-$1.0. The primordial orbital separation of the binary system should be wide enough for primary to evolve into an AGB star (wider than $2300\,R_{\odot}$). The primary loses a lot of material by the wind before it fills its Roche-lobe, and results in the stable Roche-lobe overflow (RLOF) which occurs later. After the stage of RLOF, the binary system becomes a CO WD + main sequence (MS) star system. At this stage, the secondary is still a MS star which is more massive than the primordial primary. The WD + MS system continues to evolve, and then the MS secondary becomes an AGB star and fills its Roche-lobe. After this stage, the system enters a CE phase owing to the deep convective envelope of the AGB star and the large mass ratio. In the subsequent stage, if the CE cannot be ejected, the WD will merge with the core of the AGB star during the CE phase (see also Soker 2013b).

In this article, Hurley's rapid binary volution code was adopted in our BPS approach. In this code, several processes are taken into consideration in the mass transfer process via RLOF, e.g., dynamical mass transfer, nuclear mass transfer and thermal mass transfer etc. In addition, wind accretion is also taken into consideration in this code. For details see Sections 2.1 and 2.6 in Hurley et al. (2002).

We used the standard energy equations (e.g., Webbink 1984) to calculate the output of the CE phase. The CE is ejected if
\begin{equation}
 \alpha_{\rm ce} \left( {G M_{\rm don}^{\rm f} M_{\rm acc} \over 2 a_{\rm f}}
- {G M_{\rm don}^{\rm i} M_{\rm acc} \over 2 a_{\rm i}} \right) = {G
M_{\rm don}^{\rm i} M_{\rm env} \over \lambda R_{\rm don}},
\end{equation}
where $\lambda$ is a structural parameter that depends on the evolutionary stage of the donor, $M_{\rm don}$ is the mass of the
donor, $M_{\rm acc}$ is the mass of the accretor, $a$ is the orbital separation, $M_{\rm env}$ is the mass of the donor's envelope, $R_{\rm don}$ is the radius of the donor, and the indices ${\rm i}$ and ${\rm f}$ denote the initial and final values, respectively. The right side of the equation represents the binding energy of the CE, the left side shows the difference between the final and initial orbital energy, and $\alpha_{\rm ce}$ is the CE ejection efficiency. In principle, we expect $0<\alpha_{\rm ce}\leq1$, but we often find that $\alpha_{\rm ce}$ exceeds 1 for the purpose of explaining the observed binaries (Han et al. 1995; Webbink 2008). There are two highly uncertain parameters (i.e., $\lambda$ and $\alpha_{\rm ce}$), and we combine $\alpha_{\rm ce}$ and $\lambda$ into a single free parameter $\alpha_{\rm ce}\lambda$ in this paper. We change the value of $\alpha_{\rm ce}\lambda$ to examine its influence on the birthrates and delay time of SNe Ia, and set it to be 0.01, 0.1 and 1.0 (e.g., Wang et al. 2009b).

The basic initial parameters for the Monte Carlo BPS simulations are as follows:

 (1) A constant star formation rate (SFR) of 5 ${M}_{\odot}\rm yr^{-1}$ over the past 14 Gyrs is adopted, or alternatively, it is modeled as a delta function for a single instantaneous starburst (a burst producing $10^{10} {M}_{\odot}$ in stars is assumed).

 (2) The initial mass function (IMF) which proposed by Miller \& Scalo (1979) is adopted.

 (3) A constant mass-ratio distribution is taken (e.g., Goldberg \& Mazeh 1994).

 (4) A Monte Carlo method is utilized to generate the primordial binary samples. We assume that all stars are members of binary systems and that the distribution of separations is constant in $\log a$ for wide binaries, where $a$ is separation and falls off smoothly at small separation:
\begin{equation}
a\cdot n(a)=\left\{
 \begin{array}{lc}
 \alpha_{\rm sep}(a/a_{\rm 0})^{\rm m}, & a\leq a_{\rm 0},\\
\alpha_{\rm sep}, & a_{\rm 0}<a<a_{\rm 1},\\
\end{array}\right.
\end{equation}
where $\alpha_{\rm sep}\approx0.07$, $a_{\rm 0}=10\,R_{\odot}$, $a_{\rm 1}=5.75\times 10^{\rm 6}\,R_{\odot}=0.13\,{\rm pc}$ and
$m\approx1.2$ (Han et al. 1995).

 (5) A circular orbit is assumed for all binaries. The orbits of semidetached binaries are generally circularized by the tidal force on a timescale which is much smaller than the nuclear timescale. Moreover, a binary is expected to become circularized during the RLOF. As an alternative, we also consider a uniform eccentricity distribution in the range [0, 1].

 (6) A substantially revised version that presented by Tout et al. (1997) is used to treat RLOF in Hurley's rapid binary evolution code, and the stability of the mass transfer is described with the radius-mass exponent which was defined by Webbink (1985).

 (7) Metallicities were chosen to be Z=0.02.

\begin{figure}[H]
\begin{center}
\epsfig{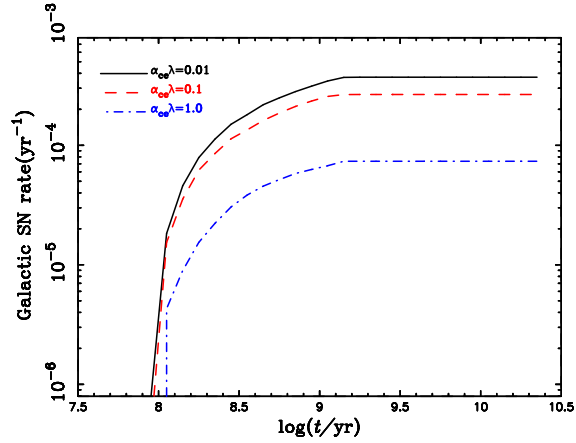} \caption{The evolution of SN Ia birthrates with time from the CD scenario for a ${\rm SFR} = 5\,M_{\odot} \rm yr^{-1})$ with different values of $\alpha _{\rm ce} \lambda$. The key to the line-styles representing different $\alpha _{\rm ce} \lambda$ is giving in the upper left corner. }
\end{center}
\end{figure}

\section{Results}

\begin{figure}[H]
\begin{center}
\epsfig{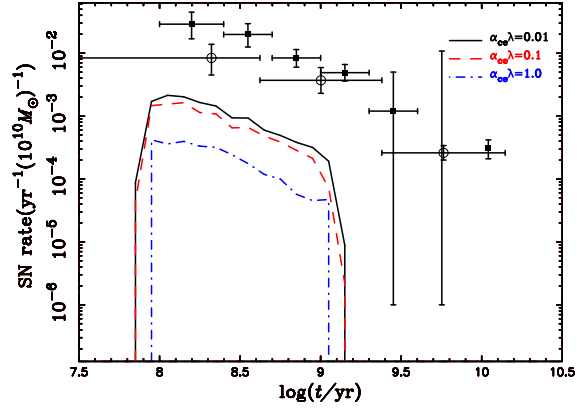} \caption{The evolution of SN Ia birthrate with time from the CD scenario for a single starburst of $10^{\rm 10}M_{\odot}$. In this figure, the spin-down time is not included in the delay time. The open circles and filled squares are taken from Maoz et al. (2011) and Totani et al. (2008), respectively. The key to the line-styles representing different $\alpha _{\rm ce} \lambda$ is giving in the upper right corner. }
\end{center}
\end{figure}

We performed three sets of simulations to systematically investigate Galactic birthrate of SNe Ia for the CD scenario by changing the model parameter to examine their influences on the final results. The Galactic SN Ia birthrate under the assumptions made here from the CD scenario is in the range of $\sim$7.4$\times 10^{-5} \rm yr^{-1}$ $-$3.7$\times 10^{-4} \rm yr^{-1}$ (see Fig. 1), which accounts for $\sim$2$-$10\% of the Galactic SN Ia birthrate observed ($\sim$3$-$4 $\times 10^{-3}\rm yr^{-1}$; Cappellaro \& Turatto 1997; Li et al. 2011). The birthrate in this paper is much lower than the value obtained from observations. In Fig. 1, we can see that the SN Ia birthrate of the CD scenario increased as the $\alpha _{\rm ce} \lambda$ is decreased. This trend can be understood as follows: the final orbital separation should be smaller to insert the same amount of energy to the envelope, which will lead to more WD$+$AGB core mergers, and hence more SNe Ia. Fig. 2 shows the evolution of SN Ia birthrates with time from the CD scenario for a single starburst with a total mass of $10^{\rm 10}M_{\odot}$, and the SNe Ia occur with no appreciable delay after merging where the spin-down time is assumed to be negligible. From this figure, we can see that SNe Ia from CD scenario occur between 70\,Myrs and 1400\,Myrs after the starburst, which means that the CD scenario might explain some young SNe Ia with short delay times. In this figure, the shortest SN Ia delay time would be mainly decided by the lifetime of a MS star with $6.5\,M_{\odot}$.

\begin{figure}[H]
\begin{center}
\epsfig{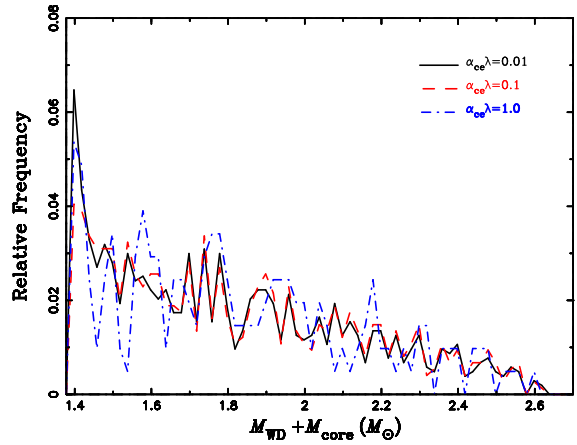} \caption{Distribution of the combined masses of the ${M}_{\rm WD}$ + ${M}_{\rm core}$ in  WD$+$AGB systems that can produced SNe Ia. Only systems with ${M}_{\rm WD} \leqslant {M}_{\rm core}$ are included in this figure. The solid, dashed and dot-dashed lines show the cases of $\alpha_{\rm ce} \lambda = 0.01$, $\alpha_{\rm ce} \lambda = 0.1$ and $\alpha_{\rm ce} \lambda = 1.0$, respectively.}
\end{center}
\end{figure}

\begin{figure}[H]
\begin{center}
\epsfig{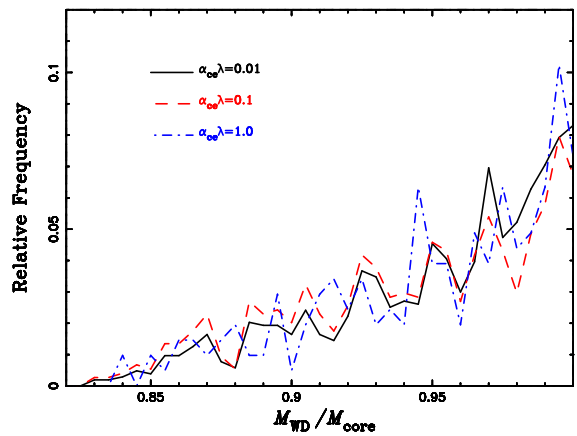} \caption{Similar to Fig. 3, but for the distribution of the mass ratio between the WD and the AGB core, ${M}_{\rm WD}/M_{\rm core}$, in the WD$+$AGB systems. }
\end{center}
\end{figure}

\begin{figure}[H]
\begin{center}
\epsfig{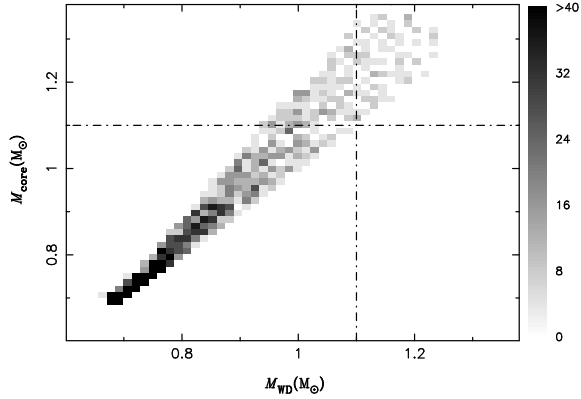} \caption{Density distribution in the initial mass plane of ${M}_{\rm WD}$ and ${M}_{\rm core}$ in WD$+$AGB systems that can produced SNe Ia, in which  we set $\alpha_{\rm ce} \lambda = 0.01$.}
\end{center}
\end{figure}

In Fig. 3, we display the mass distribution of the total mass ${M}_{\rm total}={M}_{\rm WD}+{M}_{\rm core}$ of the WD$+$AGB  systems that can ultimately produce SNe Ia. From this figure, we can see that there is a peak of ${M}_{\rm WD}+{M}_{\rm core}$ in the vicinity of $1.4{M}_{\odot}$. This trend can be understood by the initial mass function (IMF) of stars (e.g., Miller \& Scalo 1979). Fig. 4 displays the distribution of ${M}_{\rm WD}/{M}_{\rm core}$ with different $\alpha_{\rm ce}\lambda$. From this figure, we can see that there is a peak of ${M}_{\rm WD}/{M}_{\rm core}$ in the vicinity of 1.0, and almost all the values of ${M}_{\rm WD}/{M}_{\rm core}$ are above 0.8. This result is in disagreement with that of Ilkov \& Soker (2013), in which the peak of ${M}_{\rm WD}/{M}_{\rm core}$ is in the vicinity of 0.8 indicating larger AGB cores. This might be a result of the much higher mass transfer parameter adopted by Ilkov \& Soker (2013), and consequently arrived at a much lower mass ratio ${M}_{\rm WD}/{M}_{\rm core}$ (for details see Section 4).

Fig. 5 presents the density distribution in the initial mass plane of ${M}_{\rm WD}$ and ${M}_{\rm core}$ where ${M}_{\rm WD} \leqslant {M}_{\rm core}$. Note that the density distribution is predominantly concentrated in the vicinity of the diagonal in this figure. This result is in disagreement with that of Ilkov \& Soker (2013), which have a larger distribution area above diagonal indicating more massive AGB cores. For the same reason, this might be a result of the much higher mass transfer parameter adopted by Ilkov \& Soker (2013).

\begin{figure}[H]
\begin{center}
\epsfig{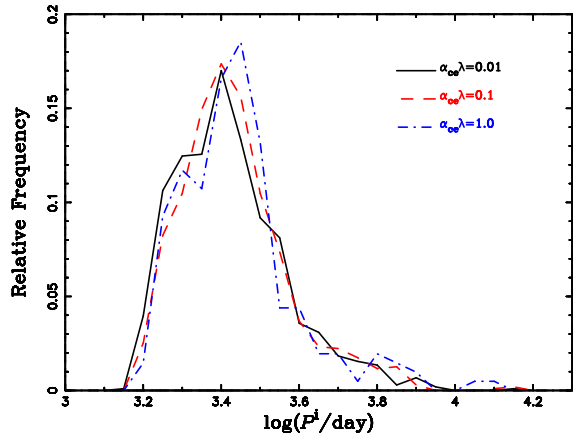} \caption{Initial orbital period distribution of WD$+$AGB systems with different values of $\alpha_{\rm ce} \lambda$, in which these WD$+$AGB systems can produced SNe Ia. }
\end{center}
\end{figure}

\begin{figure}[H]
\begin{center}
\epsfig{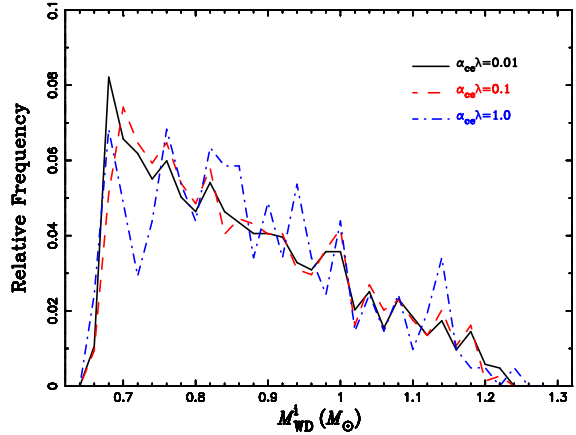} \caption{Similar to Fig. 6, but for the initial mass distribution of the WDs in WD$+$AGB systems.}
\end{center}
\end{figure}

According to our BPS approach, we also present some properties of WD$+$AGB systems that can produced SNe Ia, which would be helpful to search for potential progenitor systems of SNe Ia. Fig. 6 shows the distribution of the initial orbital periods of the WD$+$AGB systems for producing SNe Ia. We note that there is a peak at $\sim$ $10^{3.4}\,\rm d$. In Fig.7, we display the distribution of the initial masses of the CO WDs in WD$+$AGB systems. From this figure, we can see that the distribution  is in the range of $0.64-1.26\,{M}_{\odot}$, and that most of the masses are concentrated in the interval between $\sim$ $0.66\,{M}_{\odot}$ and $1.0\,{M}_{\odot}$. In Fig. 8,  we show the distribution of the initial masses of the AGB stars in WD$+$AGB systems. From this figure, we can see that the distribution range is $1.4-7.2\,{M}_{\odot}$, and that most of the masses are concentrated in the interval between $\sim$ $2\,{M}_{\odot}$ and $4.5\,{M}_{\odot}$. These properties will be helpful to constrain the progenitor scenario studied in this paper.

In Fig. 9, we present the distribution of the orbital period of surviving WD$+$WD binaries which are resulted from WD$+$AGB systems for different $\alpha_{\rm ce} \lambda$ with a uniform eccentricity distribution. From this figure, we can see that as $\alpha_{\rm ce} \lambda$ increases, the orbital period of the surviving WD$+$WD binaries also increases. This trend can be understood as follows: a lower $\alpha _{\rm ce} \lambda$ causes more WD$+$AGB core mergers and less surviving WD$+$WD binaries. Meanwhile, a lower $\alpha _{\rm ce} \lambda$ leads to closer surviving WD$+$WD binaries, the reason of which is the same as that of the afore mentioned birthrate trend. Moreover, the ratios of the number of surviving WD$+$WD systems in this figure to those that merged WD$+$AGB systems in the CD scenario for $\alpha_{\rm ce} \lambda=0.01$, $0.1$ and $1.0$ are 2.07, 3.45 and 15.54 according to our simulations, respectively. In reality, only a small portion of these surviving close WD$+$WD systems would merge by gravitational waves, as in the DD scenario, or by tidal forces at later evolution. If we take into consideration only those surviving WD+WD systems which merge within a Hubble timescale (the evolutionary timescale plus gravitational wave timescale is less than the Hubble timescale), the ratios are 1.66, 0.75 and 0.0 for $\alpha_{\rm ce} \lambda=0.01$, $0.1$ and $1.0$, respectively. From this figure, we also see that there are many surviving WD$+$WD systems, which is emerged from CE ejections, with period shorter than $0.01$ day for $\alpha _{\rm ce} \lambda$=0.01. The gravitational wave radiation will bring these surviving WD$+$WD systems to merge soon after CE ejection, and this is the another sub-channel of CD scenario (see Meng \& Yang 2012). For observational and theoretical papers on WD$+$WD binaries, see, e.g., Badenes \& Maoz (2012) and Toonen et al. (2012).

\begin{figure}[H]
\begin{center}
\epsfig{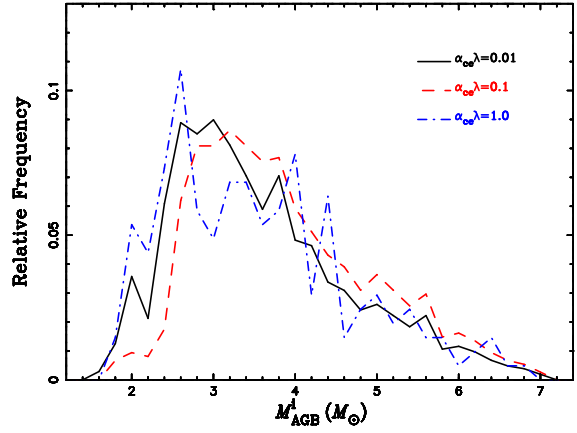} \caption{Similar to Fig. 6, but for the initial mass distribution of the AGB stars in WD$+$AGB systems. }
\end{center}
\end{figure}

\begin{figure}[H]
\begin{center}
\epsfig{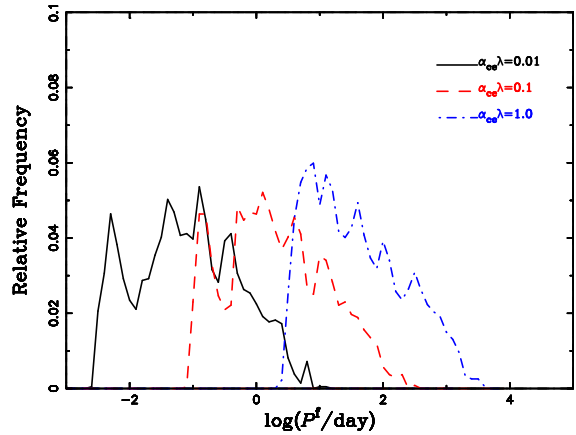} \caption{The final orbital period distribution for the surviving WD$+$WD binaries which are resulted from WD$+$AGB systems with an uniform eccentricity distribution. The solid, dashed and dot-dashed lines show the cases of $\alpha_{\rm ce} \lambda = 0.01$, $\alpha_{\rm ce} \lambda = 0.1$ and $\alpha_{\rm ce} \lambda = 1.0$, respectively. The ratios of the number of surviving systems in this figure to those that merged in the CD scenario for $\alpha_{\rm ce} \lambda=0.01$, $0.1$ and $1.0$ are 2.07, 3.45 and 15.54, respectively. If only those surviving WD+WD systems which merge within a Hubble timescale are taken into consideration, the ratios are 1.66, 0.75 and 0.0 for $\alpha_{\rm ce} \lambda=0.01$, $0.1$ and $1.0$, respectively. }
\end{center}
\end{figure}

\begin{figure}[H]
\begin{center}
\epsfig{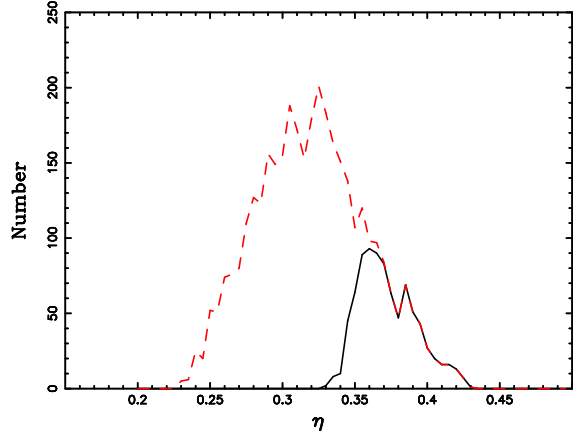} \caption{The distribution of mass transfer parameter $\eta$ which is adopted in our code, in which  we set $\alpha_{\rm ce} \lambda = 0.01$. The dashed line shows the distribution of $\eta$ for all potential WD$+$AGB systems. The solid line represents the distribution of $\eta$ for those WD$+$AGB systems which can result in SNe Ia.}
\end{center}
\end{figure}

\section{Discussion and conclusions} \label{6. DISCUSSION}

We found that the Galactic SN Ia birthrate of the CD scenario is in the range of $\sim$ $7.4\times 10^{-5}\,{\rm  yr}^{-1}$$-$$3.7\times 10^{-4}\,{\rm  yr}^{-1}$, which accounts for about $\sim$2$-$10\% of the observed value. Meng \& Yang (2012) recently obtained the SN Ia birthrate via a sub-channel by which the CD scenario can form SNe Ia when the merging process of double WDs occurs within about $10^{5}$\,yr after the CE phase, the contribution from this sub-channel of CD scenario to all SNe Ia is only about 0.1\%. Even when the birthrate of this sub-channel of the CD scenario is added, the total contribution of the CD scenario is still less than 10\%. In other words, under our assumptions the birthrate of SNe Ia due to the CD scenario is only a small fraction of the total observed SN Ia birthrate compared with the results of Ilkov \& Soker (2013). In contrast, Ilkov \& Soker (2013) claimed that the CD scenario plays a leading role in SN Ia formation, and that the birthrate induced by the CD scenario can match the observed birthrate of SNe Ia by adopted favorite values.

Obviously, our results significantly different from those of Ilkov \& Soker (2013). The main difference between our work and that of Ilkov \& Soker (2013) is the treatment of the mass transfer between the binary constituents. Ilkov \& Soker (2013) calculated the new mass of the MS secondary (${M}_{\rm 2new}$) with the formula ${M}_{\rm 2new}={M}_{2}+\eta ({M}_{1}-{M}_{\rm WD})$. At this step, the primary (with initial mass ${M}_{1}$) has evolved through the AGB phase and turned into a WD (with mass ${M}_{\rm WD}$) and the secondary (with mass ${M}_{\rm 2new}$) is still a MS star ( with initial mass ${M}_{2}$), but has accreted mass from the primary star. In order to conduct a quantitative comparison with the results of Ilkov \& Soker (2013), we take into account the distribution of the value of the mass transfer parameter $\eta$ established by our calculations (see Fig. 10). In Fig. 10, we can see that the values of the mass transfer parameter $\eta$ are in the range of $\sim$0.22$-$0.43 for all potential WD$+$AGB systems, and $\sim$0.33$-$0.43 for those WD$+$AGB systems which can result in SNe Ia. The maximum value of $\eta$ is under 0.45, which is only half the value of $\eta$ taken by Ilkon \& Soker (2013), $\eta$=0.8$-$0.9. From this figure, we can also see that the mass transfer process play a crucial role in forming SNe Ia in the CD scenario.

As a consequence of the high mass transfer parameter, the calculations of Ilkon \& Soker (2013) resulted in more massive AGB stars and larger cores (this issue has been discussed above when analysing Figs 4-5), and consequently reached a much higher birthrate of SNe Ia. Besides, Ilkov \& Soker (2013) estimated the number of SNe Ia from the CD scenario with a simple population synthesis method, and calculated the birthrate based on an ideal initial parameter space. In fact, the true initial parameter space (capable of generating WD$+$AGB systems which can result in SNe Ia) may be relatively small, and the resultant birthrate of SNe Ia will not be so high.

Observations of several over-luminous SNe Ia imply that they come from the WD explosion where the WD has a mass exceeding  the standard Chandrasekhar limit (e.g., Howell et al. 2006; Astier et al. 2006; Hicken et al. 2007; Yamanaka et al. 2010; Tanaka et al. 2010; Yuan et al. 2010; Scalzo et al. 2010). Tout et al. (2008) claimed that the formation of massive rotating WDs with strong magnetic fields might be attributed to the merger of a WD with the core of an AGB star. In Fig. 3, we can see that most of the masses of the WD$+$AGB core was larger than 1.4${M}_{\odot}$. Therefore, the CD scenario might be in a position to account for the formation of over-luminous SNe Ia, and the CD scenario contribution to SN Ia birthrates might mainly manifest itself in the form of over-luminous SNe Ia. Note that the SD model of SNe Ia may also produce over-luminous SNe Ia if the WDs have been prevented from exploding by the effect of differential rotation (e.g., Yoon \& Langer 2005; Chen \& Li 2009; Hachisu et al. 2012; Wang et al. 2014a).

Pakmor et al. (2010) proposed that some SNe Ia are the result of a violent merger of two equally massive WDs, in which the mass ratio is between 0.8 and 1.0. The main SN Ia forming mechanism of the CD scenario may be the violent prompt merger of a WD with the core of a AGB star because the mass range and the mass ratio of the WD and the AGB core meet the criteria which was proposed by Pakmor et al. (2010). For example, from Fig. 4, we can see that almost all the values of $M_{\rm WD}/M_{\rm core}$ are concentrated in the interval of 0.8$-$1.0.

In this paper, we obtained an upper limit on the birthrate of SNe Ia based on the CD scenario by taking into account all the potential WD$+$AGB systems. Under our assumptions and parameters the total contribution from all the potential WD$+$AGB systems to all SNe Ia is no more than 10\%. We note that some of our assumptions are not in consensus with others (Ilkov \& Soker 2013). We also obtained the delay time distribution of SNe Ia arising from the CD scenario, the slope of which follows a power law of $t^{-1}$ (e.g., Graur et al. 2011; Maoz et al. 2011), and the CD scenario may explain some young SNe Ia with short delay times. The young SNe Ia may play an important role in galactic chemical evolution (Scannapieco \& Bildsten 2005; Aubourg et al. 2008; Mannucci et al. 2008; Wang et al. 2009b), as a large amounts of iron would be returned to the interstellar medium much earlier than previously studies.

\begin{acknowledgements}
We acknowledge the anonymous referee for valuable comments that helped us improve the paper. We also thank Prof. Zhanwen Han for his helpful discussions and Yan Gao for his kind help to improve the language of this paper. This work is partly supported by the National Basic Research Program of China (973 programmes, 2014CB845700), the National Natural Science Foundation of China (Grant Nos. 11322327, 11390371, 11473063 and 11033008), the Foundation of State Ethnic Affairs Commission (Grant No. 12YNZ008), the Science Foundation of Key Laboratory in Software Engineering (Grant No. 2012SE402) and the Natural Science Foundation of Yunnan Province (Grant Nos. 2013FB083 and 2013HB097).

\end{acknowledgements}

\label{lastpage}


\begin{thebibliography}{}\label{thebibliography}
\bibitem[Ablimit et al. (2014)]{abl14}           Ablimit, I., Xu, X.-J., \& Li, X.-D., 2014, ApJ, 780, 80
\bibitem[Astier et al. {2006}]{Ast06}            Astier, P., Guy, J., Regnault, N., et al., 2006, A\&A, 447, 31
\bibitem[Aubourg et al. (2008)]{Aub08}           Aubourg, E., Tojeiro, R., Jimenez, R., et al., 2008, A\&A 492, 631
\bibitem[Badenes \& Maoz (2012)]{Bad12}          Badenes, C., \& Maoz, D., 2012, ApJL, 749, L11
\bibitem[Bogomazov \& Tutukov(2011)]{Bog11}      Bogomazov, A. I., \& Tutukov, A. V., 2011, Astronomy Reports, Vol.55, No.6, pp.497¨C504
\bibitem[Briggs et al. (2015)]{Bri15}            Briggs, G. P., Ferrario, L., Tout, C. A., et al., 2015, MNRAS, 447, 1713
\bibitem[Cappelllaro (1997)]{Cap97}              Cappellaro, E., Turatto, M., 1997, In: Ruiz-Lapuente, P., Cannal, R., Isern, J. (Eds.), Proceedings of the NATO Advanced Study Institute, Thermonuclear Supernovae, Vol. 486. Kluwer, Dordrecht, p. 77
\bibitem[Chen \& Li (2009)]{che09}               Chen, W.-C., \& Li, X.-D., 2009, ApJ, 702, 686
\bibitem[Dilday et al. (2012)]{Dil12}            Dilday, B., Howell, D. A., Cenko, S. B., et al., 2012, Science, 337, 942
\bibitem[Goldberg (1994)]{Gol94}                 Goldberg, D.,\& Mazeh, T., 1994, A\&A, 282, 801
\bibitem[Graur (2011)]{Gra11}                    Graur, O., Poznanski, D., Maoz, D., et al., 2011, MNRAS, 417, 916
\bibitem[Greggio \& Renzini (1983)]{Gre83}       Greggio, L., Renzini, A., 1983, A\&A,118, 217
\bibitem[Hachisu et al. (1996)]{hac96}           Hachisu, I., Kato, M., \& Nomoto, K., 1996, ApJL, 470, L97
\bibitem[Hachisu et al. (2012)]{hac12}           Hachisu, I., Kato, M., Saio, H., \&  Nomoto, K., 2012, ApJ, 744, 69
\bibitem[Han \& Podsiadlowski (2004)]{han04}     Han, Z., \& Podsiadlowski, P., 2004, MNRAS, 350, 1301
\bibitem[Han et al. {1995}]{Han95}               Han, Z., Podsiadlowski, Ph., \& Eggleton, P. P., 1995, MNRAS, 272, 800
\bibitem[Hicken (2007)]{Hic07}                   Hicken, M., Garnavich, P. M., Prieto, J. L., et al., 2007, ApJ, 669, L17
\bibitem[Hillebrandt (2000)]{Hil00}              Hillebrandt, W., \& Niemeyer, J. C., 2000, ARA\&A, 38, 191
\bibitem[Howell (2011)]{How11}                   Howell, D. A. 2011, Nature Communications, 2, 350
\bibitem[Howell (2006)]{How06}                   Howell, D. A., Sullivan, M., Nugent, P. E., et al., 2006, Nature, 443, 308
\bibitem[Hoyle (1960)]{Hoy60}                    Hoyle, F., \& Fowler, W. A., 1960, ApJ, 132, 565
\bibitem[Hurley et al. (2000)]{Hur00}            Hurley, J. R., Pols, O. R., \& Tout, C. A., 2000, MNRAS, 315, 543
\bibitem[Hurley et al. (2002)]{Hur02}            Hurley, J. R., Tout, C. A., \& Pols, O. R., 2002, MNRAS, 329, 897
\bibitem[Iben \& Tutukov (1984)]{IT84}           Iben, I., \& Tutukov, A. V., 1984, ApJS, 54, 335
\bibitem[Ilkov (2012)]{Ilk12}                    Ilkov, M., \& Soker, N., 2012, MNRAS, 419, 1695
\bibitem[Ilkov (2013)]{Ilk13}                    Ilkov, M., \& Soker, N., 2013, MNRAS, 428, 579
\bibitem[Kashi (2011)]{Kas11}                    Kashi, A., \& Soker N., 2011, MNRAS, 417, 1466
\bibitem[Kushnir (2013)]{Kus13}                  Kushnir, D., Katz, B., Dong, S., Livne, E., Fern\'andez, R., 2013, ApJ, 778, L37
\bibitem[Kuznetsova (2008)]{Kuz08}               Kuznetsova, N., Barbary, K., Connolly, B., et al., 2008, ApJ, 673, 981
\bibitem[Leibundqut (2000)]{Lei00}               Leibundgut, B., 2000, A\&ARv, 10, 179
\bibitem[Li (2011)]{Li11}                        Li, W. D., Chornock, R., Leaman, J., et al., 2011, MNRAS, 412, 1473
\bibitem[Li \& van den Heuvel (1997)]{lix97}     Li, X.-D., \& van den Heuvel, E. P. J., 1997, A\&A, 322, L9
\bibitem[Livio (2003)]{Liv03}                    Livio, M., \& Riess, A., 2003, ApJ, 594, L93
\bibitem[Livne \& Arnett (1995)]{Liv95}          Livne, E., \& Arnett, D., 1995, Apj, 452, 62
\bibitem[Maguire (2013)]{Mag13}                  Maguire, K., Sullivan, M., Patat, F., et al., 2013, MNRAS, 436, 222
\bibitem[Mannucci (2008)]{Man08}                 Mannucci, F.,Maoz, D., Sharon, K., Botticella, M. T., Della Valle, M., Gal-Yam,
A., \& Panagia, N., 2008, MNRAS, 383, 1121
\bibitem[Maoz et al. (2011)]{Mao11}              Maoz, D., Mannucci, F., Li, W., et al., 2011, MNRAS, 412, 1508
\bibitem[Maoz (2014)]{mao14}                     Maoz, D., Mannucci, F., \& Nelemans, G., 2014, ARA\&A, 52, 107
\bibitem[Matteucci \& Greggio (1986)]{Mat86}     Matteucci, F., Greggio, L., 1986, A\&A, 154, 279
\bibitem[Meng (2009)]{Men09}                     Meng, X., Chen, X., \& Han, Z., 2009, MNRAS, 395, 2103
\bibitem[Meng et al. (2011)]{Men11}              Meng, X., Chen, W. C., Yang, W. M., et al., 2011, A\&A, 525, A129
\bibitem[Meng \& Yang (2012)]{Men12}             Meng, X., \& Yang, W., 2012, A\&A, 543, A137
\bibitem[Miller \& Scalo (1979)]{Mil79}          Miller, G. E., \& Scalo, J. M., 1979, ApJS, 41, 513
\bibitem[Nelemans (2005)]{Nel05}                 Nelemans, G., \& Tout, C. A., 2005, MNRAS, 356, 753
\bibitem[Nomoto (1997)]{Nom97}                   Nomoto, K., Iwamoto, K., Kishimoto, N., 1997, Science, 276, 1378
\bibitem[Nomoto (1984)]{Nom84}                   Nomoto, K., Thielemann, F.-K., \& Yokoi, K., 1984, ApJ, 286, 644
\bibitem[Pakmor et al. (2010)]{Pak10}            Pakmor, R., Kromer, M., R\'{o}opke, F. K., et al., 2010, Nature, 463, 61
\bibitem[Parthasaratht (2007)]{Par07}            Parthasarathy, M., Branch, D., Jeffery, D. J., \& Baron, E., 2007, New Astron. Rev., 51, 524
\bibitem[Perlmutter (1999)]{Per99}               Perlmutter, S., Aldering, G., Goldhaber, G., et al., 1999, ApJ, 517, 565
\bibitem[Podsiadlowski (2008)]{Pod08}            Podsiadlowski, P., Mazzali, P., Lesaffre, P., Han, Z., \& F\"orster F., 2008, New Astron. Rev., 52, 381
\bibitem[Raskin (2013)]{Ras13}                   Raskin, C., Kasen, D., 2013, ApJ, 772, 1
\bibitem[Raskin (2009)]{Ras09}                   Raskin, C., Timmes, F. X., Scannapieco, E., Diehl, S., Fryer, C., 2009, MNRAS, 399, L156
\bibitem[Riess (1998)]{Rie98}                    Riess, A. G., Filippenko, A. V., Challis, P., et al., 1998, AJ, 116, 1009
\bibitem[Riess (2007)]{Rie07}                    Riess, A., Strolger, L.-G., Casertano, S., et al., 2007, ApJ, 659, 98
\bibitem[Rosswog (2009)]{Ros09}                  Rosswog, S., Kasen, D., Guillochon, J., Ramirez-Ruiz, E., 2009, ApJ, 705, L128
\bibitem[Ruiter et al. (2013)]{Rui13}            Ruiter, A. J., Sim, S. A., Pakmor, R., et al., 2013, MNRAS, 429, 1425
\bibitem[Saio (1985)]{Sai85}                     Saio, H., \& Nomoto, K., 1985, A\&A, 150, L21
\bibitem[Scalzo (2010)]{Sca10}                   Scalzo, R. A., Aldering, G., Antilogus, P., et al., 2010, ApJ, 713, 1073
\bibitem[Scannapieco (2005)]{Sca05}              Scannapieco, E., Bildsten, L., 2005, ApJ, 629, L85.
\bibitem[Schmidt (1998)]{Sch98}                  Schmidt, B. P., Suntzeff, N. B., Phillips, M. M., et al., 1998, ApJ, 507, 46
\bibitem[Shen (2011)]{She11}                     Shen, K. J., Bildsten, L., Kasen, D., \& Quataert, E., 2011, [arXiv: 1108. 4036]
\bibitem[Shen (2013)]{She13}                     Shen, K. J., Guillochon, J., Foley, R. J., 2013, ApJ, 770, L35
\bibitem[Soker (2013)]{Sok13a}                    Soker, N., 2013a, Proceedings of the International Astronomical Union, IAU Symposium, Volume 281, p. 72-75
\bibitem[Soker (2013)]{Sok13b}                   Soker, N., 2013b, New Astron., 18, 18
\bibitem[Soker et al. (2014)]{SG14}              Soker, N., Garc\'{i}a-Berro, E., et al., 2014, MNRAS, 437, L66
\bibitem[Soker et al. (2013)]{SK13}              Soker, N., Kashi, A., Garc\'{i}a-Berro, E., et al., 2013, MNRAS, 431, 1541
\bibitem[Sparks (1974)]{Spa74}                   Sparks, W. M., \& Stecher, T. P., 1974, ApJ, 188, 149
\bibitem[Tanaka (2010)]{Tan10}                   Tanaka, M., Kawabata, K. S., Yamanaka, M., et al., 2010, ApJ, 714, 1209
\bibitem[Toonen et al. (2012)]{Too12}            Toonen, S., Nelemans, G., \& Portegies Zwart, S., 2012, A\&A, 546, A70
\bibitem[Totani (2008)]{Tot08}                   Totani, T., Morokuma, T., Oda, T., Doi, M., \& Yasuda, N., 2008, PASJ, 60, 1327
\bibitem[Tout (1997)]{Tou97}                     Tout, C. A., Aarseth, S. J., Pols, O. R., \& Eggleton, P. P., 1997, MNRAS, 291, 732
\bibitem[Tout (2008)]{Tou08}                     Tout, C. A., Wickramasinghe, D. T., Liebert, J., Ferrario, L., Pringle, J. E., 2008, MNRAS, 387, 897
\bibitem[Tutukov \& Yungelson (1981)]{Tut81}     Tutukov, A. V., Yungelson, L. R., 1981, Nauchnye Informatsii, 49, 3
\bibitem[Wang \& Han (2010)]{WH10}               Wang, B., \& Han, Z., 2010, A\&A, 515, A88
\bibitem[Wang \& Han (2012)]{WH12}               Wang, B., \& Han, Z., 2012, New Astron. Rev., 56, 122
\bibitem[Wang et al. (2013)]{Wan13}              Wang, B., Justham, S., \& Han, Z. 2013a, A\&A, 559, A94
\bibitem[Wang et al. (2014a)]{wan14a}            Wang, B., Justham, S., Liu, Z., Zhang, J., Liu, D., \& Han, Z., 2014a, MNRAS, 445, 2340
\bibitem[Wang et al. (2014b)]{wan14b}            Wang, B., Meng, X., Liu, D., Liu, Z., \& Han, Z., 2014b, ApJL, 794, L28
\bibitem[Wang et al. (2010)]{Wan10}              Wang, B., Li, X. D., \& Han, Z., 2010, MNRAS, 401, 2729
\bibitem[Wang et al. (2009a)]{wan09a}            Wang, B., Meng, X., Chen, X., \& Han, Z., 2009a, MNRAS, 395, 847
\bibitem[Wang et al. (2009b)]{wan09b}            Wang, B., Chen, X., Meng, X., \& Han, Z., 2009b, ApJ, 701, 1540
\bibitem[Wang et al. (2013)]{wxf13}              Wang, X.-F., Wang, L., Filippenko, A. V., Zhang, T., \& Zhao, X., 2013b, Science, 340, 170
\bibitem[Webbink (1984)]{Web84}                  Webbink, R., 1984, ApJ, 277, 355
\bibitem[Webbink (1985)]{Web85}                  Webbink, R., 1985, in Interacting Binary Stars, eds. Pringle, J. E., \&  Wade, R. A. (Cambridge University Press), 39
\bibitem[Webbink (2008)]{Web08}                  Webbink, R., 2008, in Short-Period Binary Stars: Observations, Analyses, and
Results, eds. E. F. Milone, et al. (Berlin, Germany: Springer), 233
\bibitem[Whelan (1973)]{Whe73}                   Whelan, J., \& Iben, I., 1973, ApJ, 186, 1007
\bibitem[Willson (2007)]{Wil07}                  Willson, L. A., 2007, Why Galaxies Care About AGB Stars: Their Importance as Actors and Probes, 378, 211
\bibitem[Woosley \& Weaver (1994)]{Woo94}        Woosley, S. E., \& Weaver, T. A., 1994, Apg, 423, 371
\bibitem[Yamanaka (2010)]{Yam10}                 Yamanaka, M., Kawabata, K. S., Kinugasa, K., et al., 2010, ApJ, 707, L118
\bibitem[Yoon \& Langer (2005)]{YL05}            Yoon, S. C., \& Langer, N., 2005, A\&A, 435, 967
\bibitem[Yuan (2010)]{Yua10}                     Yuan, F., Quimby, R. M., Wheeler, J. C., et al., 2010, ApJ, 715, 1338

\end{thebibliography}
\end{document}